\documentclass[prl,twocolumn,superscriptaddress,floatfix,nobalancelastpage,showpacs]{revtex4}

\def\be{\begin{equation}}
\def\ee{\end{equation}}

\usepackage{amsmath}
\usepackage{amssymb}
\usepackage{graphicx}
\usepackage{dcolumn}

\begin{document}

\title{A common magnetic origin for the Invar effects in fcc iron-based ferromagnets}

\author{F. Liot}
\affiliation{Department for Computational Materials Design, Max-Planck-Institut f{\"u}r Eisenforschung GmbH, 40237 D{\"u}sseldorf, Germany}
\affiliation{Department of Physics, Chemistry and Biology (IFM), Link{\"o}ping University, SE-581 83 Link{\"o}ping, Sweden}

\author{C. A. Hooley}
\affiliation{Scottish Universities Physics Alliance (SUPA), School of Physics and Astronomy, University of St Andrews, North Haugh, St Andrews, Fife KY16 9SS, U.K.}

\begin{abstract}
Using first-principles calculations, in conjunction with Ising magnetism, we undertake a theoretical study to elucidate the origin of the experimentally observed Invar effects in disordered fcc iron-based ferromagnets. First, we show that our theory can account for the Invar effects in iron-nickel alloys, the anomalies being driven by the magnetic contributions to the average free energies. Second, we present evidence indicating that the relationship between thermal expansion and magnetism is essentially the same in all the studied alloys, including those which display the Invar effect and those which do not. Hence we propose that magnetism plays a crucial role in determining whether a system exhibits normal thermal expansion, the Invar effect, or something else. The crucial determining factor is the rate at which the relative orientation of the local magnetic moments of nearest-neighbor iron atoms fluctuates as the system is heated.
\end{abstract}

\pacs{65.40.De, 71.15.Mb, 75.10.Hk, 75.50.Bb}

\maketitle

{\it Introduction.}  Iron-based materials are used in a variety of technological applications such as springs in watches, car bodies, magnetic sensors, and heads of hard drives. Despite their ubiquity in everyday life, they exhibit intriguing phenomena ranging from high-temperature superconductivity in iron pnictides \cite{kamihara08} and Fermi-liquid breakdown in iron-niobium \cite{brando08} to the Invar effect in transition-metal alloys \cite{wassermann90}. Discovered more than 100 years ago, Invar iron-based materials display anomalous thermal expansion over broad temperature ranges. Fe$_{0.65}$Ni$_{0.35}$ was among the first in this series of substances to be discovered \cite{guillaume97}. Subsequently, a number of systems were reported, some showing ferromagnetism (e.g. Fe$_{0.68}$Pd$_{0.32}$ \cite{matsui80}) and some antiferromagnetism (e.g. Fe$_{2}$Ti \cite{wassermann98}).
\begin{figure}
\includegraphics[width=8cm]{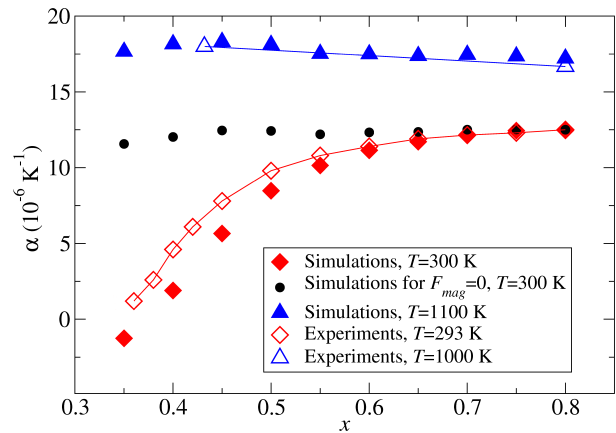}
\caption{Linear thermal expansion coefficient of an Fe$_{1-x}$Ni$_{x}$ alloy for a given temperature versus the nickel atomic concentration.  Filled diamonds show the results of our method at room temperature, and filled triangles at $T = 1100\,{\rm K}$.  The corresponding open symbols are experimental data \cite{guillaume67hayase73}.  The small filled circles show the results of our method with the magnetic contribution to the free energy removed; this underlines the crucial role of this magnetic contribution for the Invar effect.}
\label{figure1}
\end{figure}

Over the years, a paradigm has emerged in which Invar behavior in iron-based ferromagnets occurs as a result of magnetism. Despite this consensus, the nature of the mechanism giving rise to the phenomenon remains controversial. A prominent example is the debate over whether the anomaly in Fe$_{0.65}$Ni$_{0.35}$ is mainly caused by temperature-induced changes in the magnitude of local moments \cite{kakehashi81}, temperature-induced changes in the relative orientation of neighboring spins \cite{wassermann90}, partial chemical ordering \cite{crisan02}, non-collinear magnetism \cite{vanschilfgaarde99}, or something else \cite{wassermann90}. Another open question in the field is: are the anomalies observed in other systems such as Fe$_{0.72}$Pt$_{0.28}$ governed by the same essential physics as that responsible for the structural behavior in Fe$_{0.65}$Ni$_{0.35}$ \cite{wassermann90, crisan02, vanschilfgaarde99, hayn98, khmelevskyi03, ruban07}?

Obviously, a unified theory of thermal expansion in iron-based ferromagnets should capture the Invar effect in ferromagnetic disordered face-centered cubic (fcc) Fe$_{0.65}$Ni$_{0.35}$, Fe$_{0.72}$Pt$_{0.28}$, and Fe$_{0.68}$Pd$_{0.32}$ within a single framework. In principle, the linear thermal expansion coefficient, $\alpha=(1/a){(\partial a/\partial T)}_{P}$, can be derived from  the configuration-averaged Helmholtz free energy $F(a,T)$. In practice, application of density-functional theory (DFT) to \emph{ab initio} calculations of finite-temperature average free energies remains difficult, even in the adiabatic approximation where electronic, vibrational, and magnetic contributions are treated separately. Indeed, one of the major issues in implementing this strategy is how to incorporate magnetism correctly.  This difficulty may explain why only very few previous works \cite{crisan02,khmelevskyi03} which include first-principles calculations have explicitly studied thermal effects on the average length or volume of Fe-Ni, Fe-Pt, or Fe-Pd alloys.

In this Letter, we investigate theoretically thermal expansion and
its relation to magnetism in disordered fcc Fe$_{1-x}$Ni$_x$
alloys with $x$ between 0.35 and 0.8, Fe$_{0.72}$Pt$_{0.28}$, and
Fe$_{0.68}$Pd$_{0.32}$. We use an Ising model to estimate temperature-dependent
magnetic quantities such as two-spin correlation functions. On the other hand, \emph{ab initio}
total-energy calculations performed within the coherent potential approximation
(CPA) \cite{footnote3}, and the Debye-Gr{\"u}neisen model \cite{moruzzi88herper99}, provide complementary approaches for
determining average free energies.

We believe that our simulations contribute to elucidating the
essential physics of the Invar phenomenon. A key conclusion (with
broad implications) is that
the Invar effects in the Fe-Ni, Fe-Pt, and Fe-Pd alloys
may \emph{all} arise as consequences of thermal fluctuations in the relative
direction of neighboring
iron spins.

{\it Method.} To estimate the `anomalous' contribution to the thermal expansion coefficient of an Fe$_{1-x}A_{x}$ alloy as a function of temperature, we proceed as follows:

1. We calculate the average spin-spin correlation function of a nearest-neighbor Fe-Fe pair in the system at temperature $T$, $\langle S_{i}S_{j}\rangle_{FF}(T)$, and the analogous quantity for $A$-$A$ pairs, $\langle S_{i}S_{j}\rangle_{AA}(T)$. Here $S_i$ indicates whether the Ising spin on site $i$ points up ($S_i=1$) or down ($S_i=-1$). To carry out this point of the procedure, a mean-field Ising model of the M{\"u}ller-Hesse type \cite{muller83} with properly chosen exchange constants \cite{footnote1} is employed.  The
ability of an Ising model to deal with magnetism in Fe-Ni alloys
has been tested in previous work \cite{dang95}.

2. We convert the output of our Ising model into two quantities: $x_{F\uparrow}(T)$ and $x_{A\uparrow}(T)$.  Here $x_{F\uparrow}(T)$ represents the concentration of Fe atoms whose spins are up and $x_{A\uparrow}(T)$ the concentration of $A$ atoms whose spins are up, both at temperature $T$.

3. We perform {\it ab initio\/} calculations to determine the average total energy of an alloy in a ferromagnetic (FM) state or a disordered local moment (DLM) state \cite{gyorffy85, footnote2, johnson90, akai93}.  We define this state by fixing the statistics of the local moments' orientations to reproduce $x_{F\uparrow}(T)$ and $x_{A\uparrow}(T)$; their magnitudes, however, are allowed to vary.  Let us call the output of this third step $E\big(x_{F\uparrow}(T), x_{A\uparrow}(T), a\big)$, where $a$ is the average lattice spacing.

4. We add in a vibrational contribution to the free energy, estimated within the Debye-Gr{\"u}neisen model using parameters determined from step three. This yields a free energy
\begin{eqnarray}
& F\big(T, x_{F\uparrow}(T), x_{A\uparrow}(T), a\big) = & \nonumber \\
& E\big(x_{F\uparrow}(T), x_{A\uparrow}(T), a\big) + \,F_{\rm vib}(T, a). & \label{eq2}
\end{eqnarray}

5. We minimize the free energy (\ref{eq2}) with respect to $a$ to obtain the equilibrium lattice spacing at temperature $T$, $a(T)$. [$a(T)=a\big(T, x_{F\uparrow}(T), x_{A\uparrow}(T)\big)$.]

6. We evaluate the anomalous contribution to the thermal expansion coefficient, which we define as the expansion that would occur if we changed the magnetic configuration (following the Ising model), \emph{but did not otherwise heat the system}
\begin{equation}
\qquad \quad \alpha_{a}(T)= \displaystyle \lim_{\delta T \rightarrow 0} \frac{a(T') - a\big(T', x_{F\uparrow}(T), x_{A\uparrow}(T)\big)}{a(T) \,\delta T}, \label{alphaanom}
\end{equation}
%\begin{eqnarray}
%& \alpha_{a}(T)= & \nonumber \\
%& \displaystyle \lim_{\delta T \rightarrow 0}\frac{1}{a(T)} \frac{\Big[ a(T') - a\big(T', x_{F\uparrow}(T), %x_{A\uparrow}(T)\big) \Big]}{\delta T}. & \label{alphaanom}
%\end{eqnarray}
where $T' = T+\delta T$.

The counterpart to the anomalous expansion coefficient $\alpha_a$ is the normal expansion coefficient $\alpha_n$:
\be
\alpha_{n}(T)=\lim_{\delta T \rightarrow 0} \frac{a\big(T', x_{F\uparrow}(T), x_{A\uparrow}(T)\big)-a(T)}{a(T)\,\delta T}. \label{alphanorm}
\ee
This normal contribution is the expansion that would occur if we heated the system {\it without changing the configuration of the local moments\/}.
\begin{figure}
\includegraphics[width=8cm]{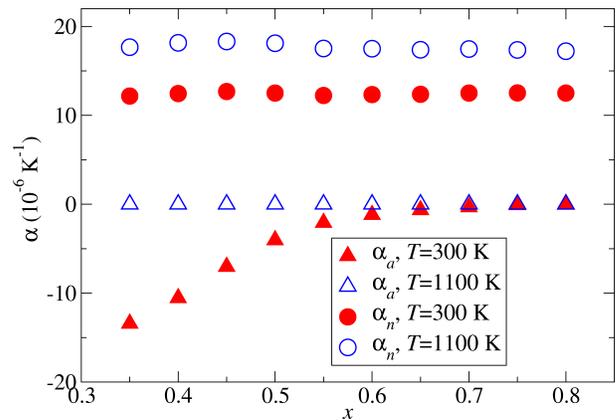}
\caption{The calculated anomalous contribution (Eq.~(\ref{alphaanom}), triangles) and the calculated normal contribution (Eq.~(\ref{alphanorm}), circles) to the thermal expansion coefficient of an Fe$_{1-x}$Ni$_{x}$ alloy for a given temperature versus the nickel atomic concentration. Filled symbols correspond to $T=300\,{\rm K}$; open symbols correspond to $T=1100\,{\rm K}$.}
\label{figure2}
\end{figure}

{\it Results and analysis.} Fig.~\ref{figure1} shows a comparison of our calculated results with measurements of the linear thermal expansion coefficient in Fe-Ni alloys \cite{guillaume67hayase73}. Our relatively simple approach reproduces the experimental data strikingly well. The figure also shows the result of running the same calculations with the magnetic contribution to the free energy $F_{mag}$ removed \cite{footnote4}. In this case there is essentially no variation of the thermal expansion coefficient with $x$, which substantiates our claim that the magnetic ingredient is crucial to the Invar effect.

The decomposition of the thermal expansion coefficient into its two parts is shown in Fig.~\ref{figure2}.
It is clear that peculiarities in thermal expansion such as the Invar effect arise from the anomalous contribution $\alpha_a$.
\begin{figure}
\includegraphics[width=8cm]{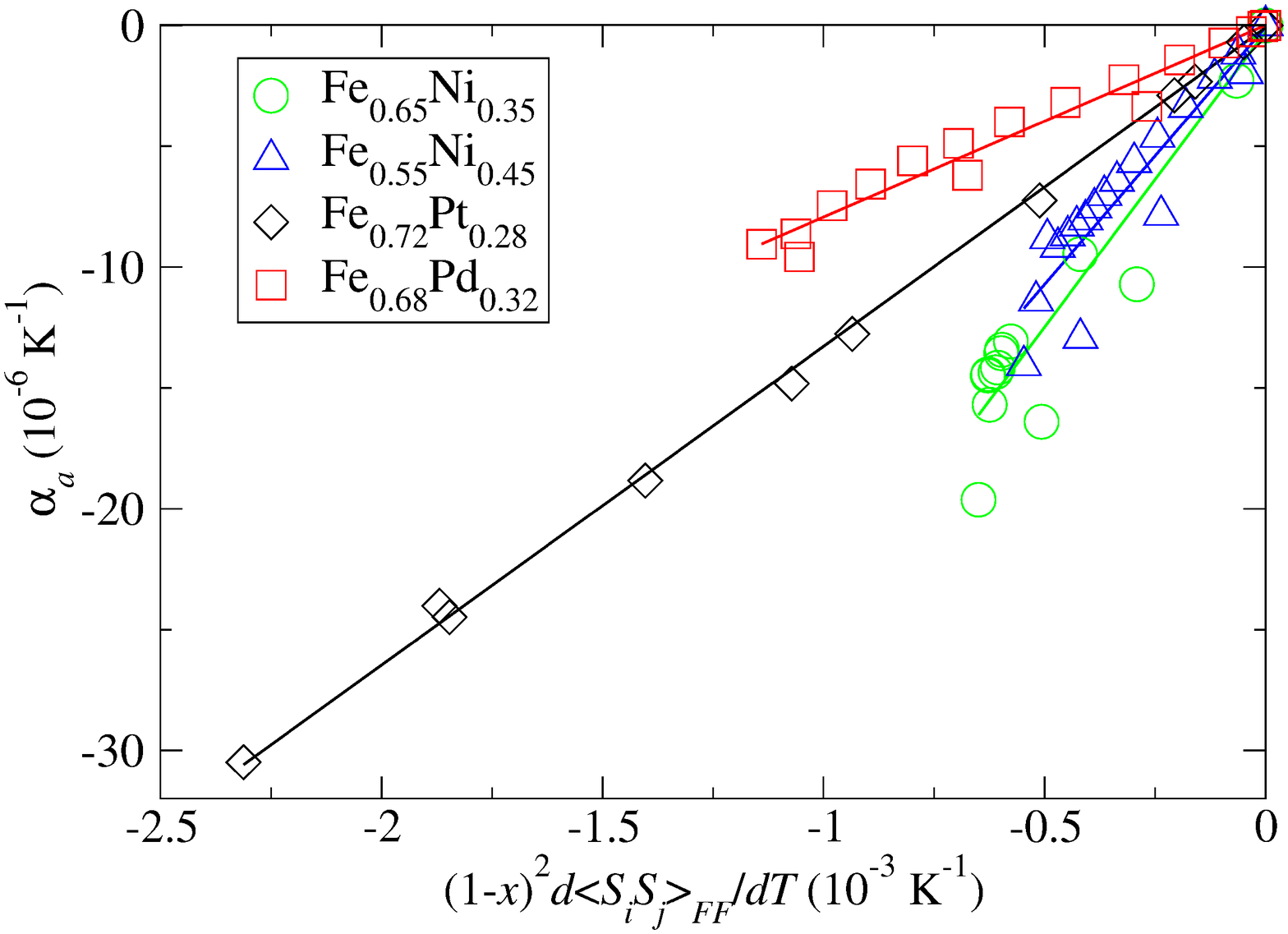}
\caption{Positive correlation between the anomalous contribution to the thermal expansion coefficient (calculated using equations (\ref{eq2}) and (\ref{alphaanom})) and the magnetic quantity $(1-x)^{2} d \langle S_{i}S_{j}\rangle_{FF}/{dT}$ (calculated using the mean-field Ising model), for various Fe-based ferromagnets.}
\label{figure3}
\end{figure}

Thus the main interpretive question to ask is: What is the magnetic origin of $\alpha_a$?
To explore this, consider our results presented in Fig.~3 for a number of Fe-based systems: (i) All of the alloys show a positive correlation between $\alpha_a$ and the product of the concentration of nearest-neighbor Fe-Fe pairs with the rate at which the average spin-spin correlation function changes with temperature \cite{footnote7},
\begin{equation}
(1-x)^{2} \frac{d \langle S_{i}S_{j}\rangle_{FF}}{dT}. \label{magquant}
\end{equation}
Basically, the larger the rate at which iron moments disorder on heating, the larger the negative deviation of thermal expansion from normal behavior. (ii) The slopes of the correlation lines are of the same order of magnitude. The existence of a connection between the thermal expansion anomaly and the temperature derivative of a two-spin correlation function is supported by previous analytical work \cite{callen65}.
\begin{figure}
\includegraphics[width=8cm]{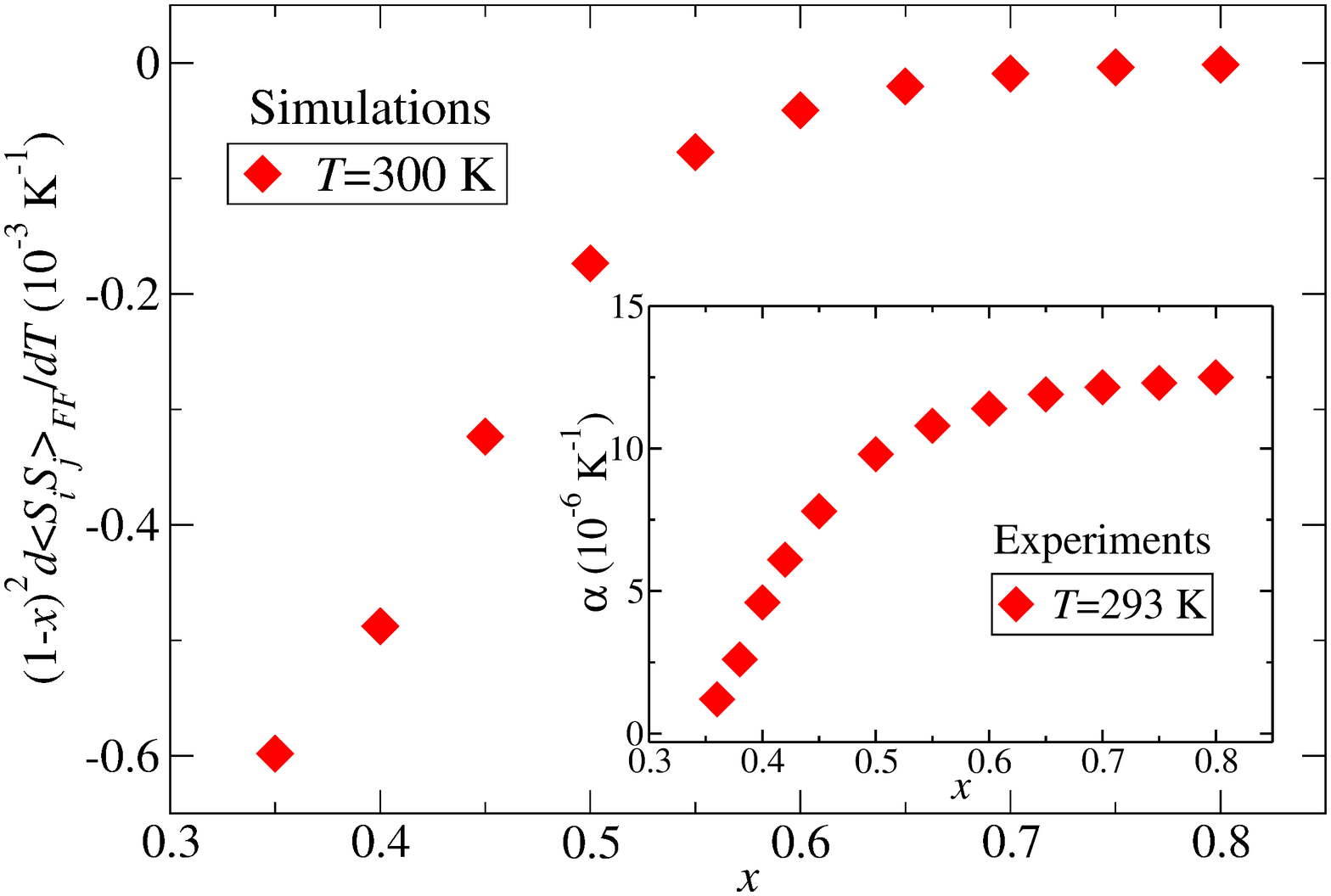}
\caption{The magnetic quantity $(1-x)^{2} d \langle S_{i}S_{j}\rangle_{FF}/{dT}$ of an Fe$_{1-x}$Ni$_x$ alloy at room temperature plotted against the nickel concentration, according to the mean-field Ising model. The inset shows an experimental result for the thermal expansion coefficient of Fe-Ni compounds.}
\label{figure4}
\end{figure}
\begin{figure}
\includegraphics[width=8cm]{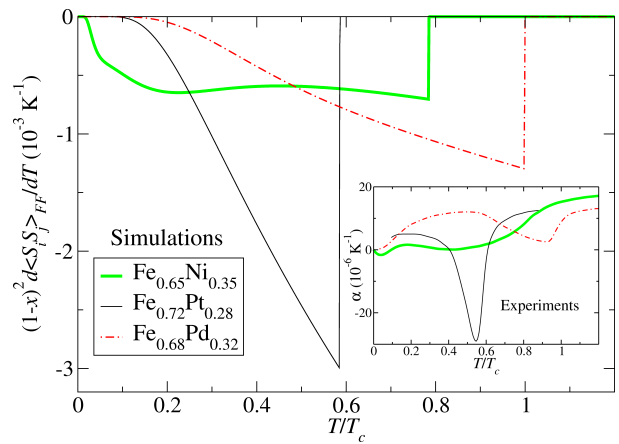}
\caption{Estimated magnetic quantity of the Invar alloys Fe$_{0.65}$Ni$_{0.35}$ (thick solid line), Fe$_{0.72}$Pt$_{0.28}$ (thin solid line), and Fe$_{0.68}$Pd$_{0.32}$ (dashed-dotted line) as a function of the reduced temperature $T/T_{c}$, where $T_c$ corresponds to the Curie temperature of Fe$_{0.68}$Pd$_{0.32}$. The inset shows the temperature dependence of the thermal expansion coefficient of the systems, as determined from experiments \cite{matsui78sumiyama79, matsui80}.  As a result of our mean-field approximation, our theoretical value for $T_c$ (782\,K) is an overestimate compared to the experimental value (637\,K) \cite{footnote1}.}
\label{figure5}
\end{figure}
 
Our analysis indicates that the thermal expansion coefficient is similarly related to the magnetic quantity (4) in \emph{all} the studied  ferromagnets (Fe$_{1-x}$Ni$_x$ with $x$ between 0.35 and 0.8, Fe$_{0.72}$Pt$_{0.28}$, and Fe$_{0.68}$Pd$_{0.32}$) \cite{footnote9}, and that the temperature evolution of the magnetic quantity (4) plays a major role in determining whether such a system exhibits normal expansion, the Invar effect, or something else. Features in the structural behavior of the systems which were observed experimentally (see the insets in Figs.~4 and~5), but had remained unexplained, can now be interpreted on the basis of the abovementioned insight and our theoretical results displayed in Figs.~4 and~5. (i) The drop in the thermal expansion coefficient of Fe$_{1-x}$Ni$_x$ at room temperature when the nickel concentration is reduced arises from the steep decrease of the magnetic term $(1-x)^{2} d \langle S_{i}S_{j}\rangle_{FF} / d T$. (ii) The fact that the expansivity of Invar Fe$_{0.72}$Pt$_{0.28}$ diminishes significantly as $T/T_{c}$ is raised from 0.2 to 0.5 whereas that of Invar Fe$_{0.65}$Ni$_{0.35}$ does not reflects the different behaviors of their magnetic quantities in this interval: $(1-x)^{2} d \langle S_{i}S_{j}\rangle_{FF} / d T$ decreases drastically in the Fe-Pt case, but remains almost constant in that of Fe-Ni.

To conclude, we point out that it would be highly desirable (i) to go beyond the present mean-field Ising model; and (ii) to investigate the effect of pressure on the temperature-dependence of the magnetic quantities \cite{liotxx}. Such investigations may lead to a unified theory for the Invar effect and the pressure-induced Invar effect \cite{dubrovinsky01winterrose09} which has been the subject of recent debate.

The interest and support of I. A. Abrikosov and J. Neugebauer are gratefully acknowledged. F. L. thanks B. Alling, M. Ekholm and P. Steneteg for helping him with calculations. This work was supported by grants from the Swedish Research Council (VR), the Swedish Foundation for Strategic Research (SSF), the G{\"o}ran Gustafsson Foundation for Research in Natural Sciences and Medicine, the EPSRC (UK), the Scottish Universities Physics Alliance, and the HPC-Europa project.

%\bibliographystyle{apsrev}
%\bibliographystyle{aps5etal}
%\bibliography{SrFe2P2Bib}

\begin{thebibliography}{31}
\bibitem{footnote3}
The first-principles calculations of average total energies are performed within the framework of the exact muffin-tin orbitals (EMTO) theory using the full charge density (FCD) technique \cite{vitos01}. The problem of substitutional chemical disorder is treated within the coherent potential approximation (CPA) \cite{vitos-abrikosov01}. The integration over the irreducible part of the Brillouin zone is done over several thousands of $\mathbf{k}$-points distributed according to the Monkhorst-Pack scheme \cite{monkhorst76}. All the calculated bulk moduli and Gr\"uneisen constants are found to be of the order of 2 Mbar, and 2. Finally, note that the computations are performed within the generalized gradient approximation (GGA) \cite{perdew96} for the exchange-correlation energy functional.
\bibitem{footnote4}
In this paper, we define $F_{mag}(T, a)$ as the difference between the energies $E\big(x_{F\uparrow}(T), x_{A\uparrow}(T), a\big)$ and $E\big(x_{F\uparrow}(0), x_{A\uparrow}(0), a\big)$.
\bibitem{footnote7}
According to previous numerical simulations \cite{liot09}, the length of an Fe-Fe nearest-neighbor bond is peculiarly sensitive to the relative orientation of the local moments on the two iron sites. Loosely speaking, as the configuration changes from $\uparrow\uparrow$ to $\uparrow\downarrow$, the bond contracts. We speculate that this effect and the correlation between $\alpha_a$ and (\ref{magquant}) are two phenomena, one local and the other global, induced by the intersite distance dependence of the magnetic coupling between Fe nearest-neighbor atoms.
\end{thebibliography}

\end{document}